\documentclass[aps,prb,reprint,amsmath,amssymb,showpacs,floatfix]{revtex4-1}

\usepackage{graphicx,units}

\newcommand{\ket}[1]{\ensuremath{\vert{#1\rangle}}} 

\newcommand{\braket}[2]{\ensuremath{{\langle #1}\vert{#2 \rangle}}}
\newcommand{\ketbra}[2]{\ensuremath{|{#1 \rangle}{\langle #2}|}}
\newcommand{\op}[1]{\hat{#1}}

\newcommand{\I}{\text{i}}
\newcommand{\E}{\text{e}}
\providecommand{\abs}[1]{\left\lvert#1\right\rvert}

\usepackage[colorlinks,citecolor=blue,urlcolor=blue,linkcolor=blue,hyperindex,breaklinks]{hyperref}

\begin{document}

\title{Delayed-choice quantum eraser for the undergraduate laboratory}
\author{James M.\ Ashby}
\author{Peter D.\ Schwarz}
\author{Maximilian Schlosshauer}
\thanks{Author to whom correspondence should be addressed. Electronic mail: \texttt{schlossh@up.edu}}
\affiliation{Department of Physics, University of Portland, 5000 North Willamette Boulevard, Portland, Oregon 97203}

\date{\today}

\begin{abstract} 
In a delayed-choice quantum eraser, interference fringes are obtained by erasing which-way information after the interfering particle has already been irreversibly detected. Following an introductory review of delayed-choice experiments and quantum erasure, we describe the experimental realization of an optical delayed-choice quantum eraser, suitable for advanced undergraduates, based on polarization-entangled pairs of single photons. In our experiment, the delay of the erasure is implemented using two different setups. The first setup employs an arrangement of mirrors to increase the optical path length of the photons carrying which-way information. In the second setup, we use fiber-optic cables to elongate the path of these photons after their passage through the polarization analyzer but prior to their arrival at the detector. We compare our results to data obtained in the absence of a delay and find excellent agreement. This shows that the timing of the erasure is irrelevant, as also predicted by quantum mechanics. The experiment can serve as a valuable pedagogical tool for conveying the fundamentals of quantum mechanics. 

\vspace{.1cm}

\noindent Journal reference: \emph{Am.\,J.\,Phys.\ }\textbf{84}, 95--105 (2016), DOI: \href{http://dx.doi.org/10.1119/1.4938151}{10.1119/1.4938151}
\end{abstract}

\maketitle

\section{Introduction}

Delayed-choice quantum erasure, inspired by a Gedanken experiment of Wheeler's\cite{Wheeler:1978:az,Wheeler:1983:ll} and first proposed by Scully and Dr{\"u}hl,\cite{Scully:1982:yb} vividly illustrates central features of quantum mechanics. When a particle passes through an interferometer, its path and the relative phase between the two possible paths are complementary observables. If the paths are in principle experimentally distinguishable through the presence of which-way information, then no interference can be observed, i.e., no phase information can be obtained. Indeed, there is a precise tradeoff between the visibility of the interference pattern and the available amount of which-way information.\cite{Wooters:1979:az,Mittelstaedt:1987:om,Greenberger:1988:oo,Jaeger:1995:um,Englert:1996:km,Durr:2000:za} For the loss of interference to occur, it does not matter whether a which-way measurement is actually carried out. It suffices that there exists the mere possibility of retrieving which-way information from a suitable future measurement: ``It is what the experimenter can do, not what he bothers to do, that is important.''\cite{Greenberger:1993:ii}

In a quantum eraser,\cite{Scully:1982:yb,Scully:1991:yb} the which-way information is encoded in ancillary degrees of freedom, typically in a second particle entangled with the first. An appropriately chosen ``erasure'' measurement is then performed on the ancilla to render the which-way information unobtainable. If the signal from the interferometer is subsequently correlated with the outcome of the erasure measurement, an interference pattern can be reconstructed. In this way, quantum erasure can be understood as a ``sorting'' or ``tagging'' of the data from the interferometer conditional on the additional information gained from the erasure measurement.\cite{Englert:1999:aq} According to quantum mechanics, the temporal order of measurements on different systems is irrelevant to the resulting statistics, even if the systems are entangled. Therefore, it does not matter when the erasure measurement is performed. In particular, we can delay the measurement long after the particle has passed through the interferometer. This protocol is known as a delayed-choice quantum eraser.

Our experiment implements a delayed-choice quantum eraser using pairs of polarization-entangled photons produced by spontaneous parametric downconversion. Our setup for generating, manipulating, and detecting single photons follows the approach developed by the Beck group at Whitman College; see Refs.~\onlinecite{Thorn:2004:za,Gogo:2005:oo,Carlson:2006:za,Beck:2007:uu,Beck:2012:az,Beck:website} for details. Specifically, the quantum-eraser part of our experiment (without delayed choice) is essentially the same as described by Gogo \emph{et al.} in Ref.~\onlinecite{Gogo:2005:oo}. Our setup, however, adds a delay stage that ensures that the erasure measurement happens only after the signal photon has passed through the interferometer and has been irreversibly measured. While from an experimental point of view this constitutes a relatively minor modification, it establishes a significantly different conceptual situation and offers the opportunity to incorporate the delayed-choice paradigm into an undergraduate experiment.

The principle of our experiment is as follows. One photon (to be referred to as the signal) passes through a polarization interferometer, such that the path through the interferometer depends on the polarization of the photon. Because of the polarization correlations between the signal photon and the entangled second photon (to be referred to as the idler), which-way information may be obtained, if only in principle, by a polarization measurement on either the signal or the idler photon, precluding the observation of an interference signal. However, by performing polarization measurements on the signal and idler photons in a rotated basis, the which-way information becomes obliterated. In this case, the coincidence counts between the signal and idler photons exhibit a sinusoidal dependence on the path length through the interferometer, i.e., an interference pattern is observed. In our experiment, the delay of the erasure measurement is implemented in two alternative ways. In the first arrangement, we introduce about two meters of additional optical distance in the idler arm. In the second arrangement, we elongate, by several meters, the fiber-optic cable transmitting the idler photons to the detector. 

The experiment described here has been set up, carried out, and analyzed by undergraduate students at our institution as a part of several student research projects. We have also incorporated it into our upper-division laboratories. Its modular structure allows it to be easily adapted to the implementation of related experiments that use correlated pairs of single photons, for example, a proof of the existence of photons,\cite{Thorn:2004:za} single-photon interference,\cite{Galvez:2005:uu,Beck:2012:az} tests of local realism,\cite{Dehlinger:2002:aa,Dehlinger:2002:tt,Carlson:2006:za} and quantum state tomography.\cite{Beck:2012:az,Dederick:2014:ll,Burch:2015:tt} The interested reader is referred to Refs.~\onlinecite{Beck:2012:az,Beck:website,Galvez:2005:uu} for details on the implementation of these and similar experiments in an undergraduate setting. The delayed-choice quantum eraser can also serve as an excellent pedagogical tool, as it incorporates and highlights central quantum-mechanical concepts such as interference, distinguishability, complementarity, measurement, information, multipartite states, entangled versus mixed states, and causality. To help undergraduate students understand and combine these concepts in a concrete context, in our Quantum Mechanics course students first perform a theoretical analysis of the quantum eraser and then carry out the experiment and analyze the data.

This paper is organized as follows. In Sec.~\ref{sec:hist-basic-conc} we review the history and basic principles of both delayed-choice interferometry and quantum erasure. In Sec.~\ref{sec:theory} we give a theoretical description of a quantum eraser based on polarization interferometry. We describe our experimental setup in Sec.~\ref{sec:experiment} and report results in Sec.~\ref{sec:results}. We discuss our findings in Sec.~\ref{sec:discussion}.

\section{\label{sec:hist-basic-conc}History and basic concepts}

\subsection{\label{sec:delayed-choice}Delayed-choice interferometry}

\begin{figure}
\includegraphics[scale=1]{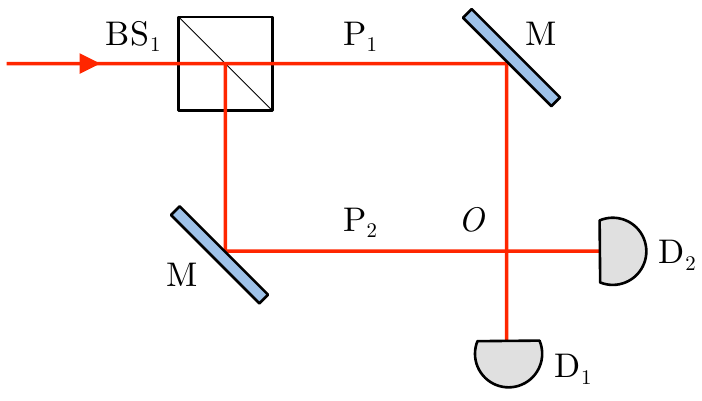}
\caption{\label{fig:wheeler}(Color online) (Color online) Wheeler's delayed-choice thought experiment based on a Mach--Zehnder interferometer. A photon passes through the 50--50 beamsplitter BS$_1$. By virtue of two mirrors (M), the paths P$_1$ and P$_2$ through the interferometer cross again at point $O$. Photons are registered by detectors D$_1$ and D$_2$. A second 50--50 beamsplitter can be inserted or removed at location $O$ to choose between two complementary measurements. In Wheeler's thought experiment, this choice is delayed until after the photon has already entered the interferometer.}
\end{figure}

While the basic problem of delayed choice was already implicitly posed in the 1930s by von Weizs\"acker\cite{Weizsacker:1931:aa,Weizsacker:1941:aa} in the context of his discussion of Heisenberg's gamma-ray microscope,\cite{Heisenberg:1927:uu} the idea of a delayed-choice implementation of an interference experiment was first stated explicitly by Wheeler.\cite{Wheeler:1978:az,Wheeler:1983:ll} Wheeler's original proposal employs the Mach--Zehnder interferometer shown in Fig.~1. Photons are incident on a  50--50 beamsplitter (denoted BS$_1$ in the figure) such that there are two possible subsequent paths, P$_1$ and P$_2$. A photon traveling down path P$_1$ (P$_2$) will pick up a phase shift $\phi_1$ ($\phi_2$), where the phase difference $\Delta \phi=\phi_2-\phi_1$ can be adjusted by changing the relative lengths of the two paths. Past the crossing point $O$ of the two paths, photon detectors D$_1$ and D$_2$ are placed as shown. Then for each photon sent through the beamsplitter, one of the two detectors will click, and both detectors have the same probability of clicking, no matter what the path lengths are. The story this observation tempts us to tell is that the photons behave like particles, each traveling down one of the two paths. On the other hand, if we insert a second beamsplitter BS$_2$ at location $O$, then the detector clicks exhibit a sinusoidal dependence on the phase difference $\Delta \phi$. In particular, for certain values of $\Delta \phi$, only one of the detectors will click for every photon entering the interferometer while the other detector will remain silent. This observation suggests an interference phenomenon involving both paths, and thus, it tempts us to associate a wave picture for the photon in which, as Wheeler put it, ``the arriving photon came by both routes.''\cite{Wheeler:1983:ll} 

Wheeler  proposed to delay the choice of whether to insert the beamsplitter BS$_2$---i.e., whether to observe particle or wave properties---until \emph{after} the photon has already passed the first beamsplitter.\cite{Wheeler:1978:az,Wheeler:1983:ll} To heighten the drama, Wheeler even considered a delayed-choice experiment on a cosmological scale, where light originating from a distant star experiences gravitational lensing by an intervening galaxy, implementing a kind of cosmic interferometer.\cite{Wheeler:1983:ll} (This move is not unlike Schr\"odinger's when he used his eponymous cat to highlight the quantum measurement problem.\cite{Schrodinger:1980:ps}) If one holds on to a naive realistic picture in which the particular physical arrangement---in our example, the absence or presence of BS$_2$---forces the photon to travel along either just one path or both paths, then one runs into a paradox, since the arrangement is only chosen once the photon is already inside the interferometer. According to Wheeler:\cite{Wheeler:1983:ll} 
\begin{quote}
Thus one decides whether the photon ``shall come by one route, or by both routes'' after it has ``\emph{already done} its travel.'' \dots\ [W]e have a strange inversion of the normal order of time. We, now, by moving the mirror [in our example, the second beamsplitter] in or out have an unavoidable effect on what we have a right to say about the \emph{already} past history of that photon.
\end{quote}
As other authors have also noted (see, e.g., Refs.~\onlinecite{Ionicioiu:2011:az,Englert:1999:aq}), such language may have contributed to some of the popular misconceptions surrounding delayed-choice quantum experiments. The problem, as we see it, is the element of retrodiction implied by expressions such as ``what we have a right to say about the already past history of that photon,'' and we will come back to such issues in Sec.~\ref{sec:notion-which-way}. It should be noted, however, that Wheeler himself tried to clarify the matter, repeatedly insisting that ``no phenomenon is a phenomenon until it is an observed phenomenon''\cite{Wheeler:1978:az} and that ``the past has no existence except as it is recorded in the present.''\cite{Wheeler:1983:ll} Interestingly, the lesson to be drawn from delayed-choice experiments was already anticipated by Bohr\cite{Bohr:1949:mk} (in his discussion of Einstein's ``photon box'' thought experiment; see pp.~225--230 of Ref.~\onlinecite{Bohr:1949:mk}) in what Wheeler\cite{Wheeler:1978:az} has called ``that solitary and pregnant sentence'':
\begin{quote}
It obviously can make no difference as regards observable effects obtainable by a definite experimental arrangement, whether our plans of constructing or handling the instruments are fixed beforehand or whether we prefer to postpone the completion of our planning until a later moment when the particle is already on its way from one instrument to another.
\end{quote}

Perhaps the closest realization of Wheeler's proposal as shown in Fig.~1 is the experiment by Jacques \emph{et al.}\cite{Jacques:2007:un} In this experiment, single photons traversed a 48-m polarization interferometer. The choice of the final polarization measurement (producing either path or phase information) was made by a quantum random number generator and was relativistically separated from the photon's entry into the interferometer. Delayed-choice experiments with quantum control have also been implemented,\cite{Ionicioiu:2011:az,Tang:2012:yb} and the delayed-choice paradigm has been applied to areas such as entanglement swapping.\cite{Peres:2000:az,Sciarrino:2002:za,Ma:2012:uu}

\subsection{\label{sec:quantum-erasure}Quantum erasure}

Consider now the situation in which, instead of directly measuring phase or path information by inserting or removing the second beamsplitter BS$_2$, we keep BS$_2$ in place at all times. While the photon passes through the interferometer, we let it appropriately interact with an ancilla in such a way that the spatial degree of freedom of the photon becomes quantum-correlated (entangled) with the ancilla. The ancilla is typically realized in the form of a spatially separated particle distinct from the interfering photon.\footnote{While not physically necessary for quantum erasure to obtain, this separation aids pedagogical clarity and is therefore one of the desirable features of an ``ideal'' quantum eraser.\cite{Kwiat:1994:za} It is also a prerequisite for delayed-choice implementations that causally disconnect the erasure measurement from the interfering particle.\cite{Ma:2013:un}} Then, even after the photon paths P$_1$ and P$_2$ are recombined at BS$_2$, they remain distinguishable in principle through the correlations with the ancilla. As a consequence, although BS$_2$ is present (corresponding to a phase measurement in Wheeler's original scenario), no interference is observed. That is, no phase information can be gained: The pattern of detector clicks will be random, just as in the case of the direct path measurement made by removing BS$_2$, and no dependence on the phase difference $\Delta \phi$ is observed. Although we have not actually measured the path of the photon, the fact that the ancilla has encoded which-way information about the particle (in a sense to be clarified in Sec.~\ref{sec:notion-which-way}) is sufficient to preclude the observation of interference. It is important to emphasize that this effect is purely a consequence of quantum correlations with the ancilla, leading to an in-principle distinguishability of the different paths through the interferometer. Thus, loss of interference does not require a physical disturbance of the photon in the sense sometimes associated with the uncertainty principle.\cite{Scully:1991:yb,Durr:2000:za,Rozema:2012:az} 

We  may, however, recover an interference pattern in the following way. Through a suitable measurement of the ancilla, we project its quantum state onto a state that represents equal probabilities of finding the photon in path P$_1$ or path P$_2$ in a subsequent measurement. In this manner, the two paths have become indistinguishable and which-way information is said to have been ``erased'':\cite{Scully:1982:yb,Scully:1991:yb} We have closed the door to the possibility of finding out anything meaningful (nonrandom) about the photon's path. At the same time, however, we have gained new information, because the process of erasure is just another measurement. As we will make precise in Sec.~\ref{sec:theory}, the outcomes of this erasure measurement provide exactly the information necessary to decompose the photon data consisting of random clicks into two out-of-phase interference patterns. We thus have a process of fine-graining, based on newly acquired information from the ancilla measurement, of the statistical data produced by the measurement on the photon. 

In a quantum eraser, then, the choice between interference and path information is not implemented by a modification of the interferometer itself (as in Wheeler's proposal), but by the choice of measurement on the ancilla. Whether we first measure the photon and then the ancilla or the other way round does not affect the joint probabilities predicted by quantum mechanics. In this way, the delayed choice open to the experimenter in the case of the quantum eraser is of a somewhat different flavor than in Wheeler's proposal. In the latter, one deals with the choice between two complementary measurements on a single photon and delays the measurement until after a point when intuition would suggest that the photon would have had to choose between two mutually exclusive histories to account for the observed results of a \emph{subsequent} measurement on the photon. In a delayed-choice quantum eraser, the measurement on the photon is fixed while the delayed action is taken on the ancilla. Because the which-way information is independently encoded in the ancilla, its erasure can be delayed even until after the signal photon has already been detected. (See Refs.~\onlinecite{Scully:1998:yy,Bahrami:2010:tt} for a detailed operational analysis of delayed-choice quantum erasure.)

The first experimental realization of a quantum eraser meeting the criteria proposed by Kwiat \emph{et al.}\cite{Kwiat:1994:za} was described by Herzog \emph{et al.}\cite{Herzog:1995:lb} The first delayed-choice quantum eraser---an optical analogue of the original proposal by Scully and Dr{\"u}hl\cite{Scully:1982:yb}---was reported by Kim \emph{et al.}\cite{Kim:2000:az} A delayed-choice quantum eraser based on a double-slit interferometer, inspired by the proposal of Scully, Englert, and Walther,\cite{Scully:1991:yb} has been realized by Walborn \emph{et al.}\cite{Walborn:2002:tz} The recent experiment by Ma \emph{et al.}\cite{Ma:2012:uu} closed the communication loophole by using a 144-km free-space separation between the interferometer and the location of the erasure measurement (see also Sec.~\ref{sec:delay-stage}). While realizations of a quantum eraser suitable for the undergraduate laboratory have been previously described by Galvez \emph{et al.}\cite{Galvez:2005:uu} and Gogo \emph{et al.},\cite{Gogo:2005:oo} these experiments did not implement a delayed choice of the erasure measurement.

We note a connection between quantum erasure and environment-induced decoherence. In a decoherence process,\cite{Zurek:2002:ii,Schlosshauer:2007:un} information distinguishing the different components in the system's superposition state is encoded, via an entangling interaction, in environmental degrees of freedom. As a consequence, interference between these components can no longer be observed by virtue of a local measurement performed on the system, just as is the case in a quantum-eraser experiment. In contrast with a quantum eraser, however, in the case of decoherence one cannot, in practice, reconstruct an interference pattern by an appropriate erasure-type measurement of the environment because of the large number of (experimentally uncontrolled) environmental degrees of freedom that have interacted with the system.

\subsection{\label{sec:notion-which-way}The notion of which-way information}

Before moving on, we feel compelled to clarify the notion of which-way information, since it is so central to discussions of the quantum eraser. Let us formalize matters by denoting the spatial states of the photon corresponding to the paths P$_1$ and P$_2$ through the interferometer by $\psi_1(x)$ and $\psi_2(x)$. The interaction with the ancilla is such that the state of the ancilla evolves into $\ket{1}$ when the photon is sent along path P$_1$ only, and it evolves into $\ket{2}$ when the photon is sent along path P$_2$ only, where we take $\ket{1}$ and $\ket{2}$ to be orthogonal (i.e., perfectly distinguishable). Since BS$_1$ produces a superposition of $\psi_1(x)$ and $\psi_2(x)$, it follows from the linearity of the time evolution that the final composite photon--ancilla state right before the photon reaches BS$_2$ is the entangled state 
\begin{equation}\label{eq:hvb}
\ket{\Psi(x)}=\frac{1}{\sqrt{2}}\left(\psi_1(x) \ket{1}+\E^{\I\Delta \phi}\psi_2(x) \ket{2}\right).
\end{equation}
If we were to measure the ancilla in the $\{ \ket{1},\ket{2} \}$ basis and find, say, $\ket{1}$, we could immediately infer that the photon must now be in the quantum state $\psi_1(x)$; this is an experimentally verifiable correlation. Here, the expression ``must now be'' simply means that we can predict with certainty (to echo the famous phrase of Einstein \emph{et al.}\cite{Einstein:1935:dr}) that a subsequent measurement of an observable that has $\psi_1(x)$ as one of its possible outcomes will give the result $\psi_1(x)$ with a probability of 1. More precisely, the photon now ``belongs to a subensemble whose statistical properties are correctly accounted for by $\psi_1(x)$.''\cite{Englert:1999:aq} It does \emph{not} mean that the photon possessed a definite path \emph{prior} to the measurement of the ancilla. In fact, that such a view is untenable is, we think, precisely the upshot of the experimentally observed violations of Bell's inequalities. Quantum states encode probabilities of \emph{future} measurements, and for an entangled state such as \eqref{eq:hvb}, we must not make retrodictive statements about one system based on the measurement on the other system (as EPR's ``criterion of reality'' attempted to do\cite{Einstein:1935:dr}). 

Thus, when we say that the ancilla ``encodes which-way information'' about the photon via the state~\eqref{eq:hvb}, this must not be understood as information about a definite path of the photon, for the question of path simply cannot have a definite answer at this point. Rather, it means that there is a procedure that allows us to \emph{distinguish}, in principle, the two paths, in the precise operational sense that measuring the ancilla and finding the outcome $\ket{1}$ will allow us to conclude that a subsequent measurement of the photon will be more likely to find, say, the outcome $\psi_1(x)$ rather than $\psi_2(x)$. If we can predict with certainty (probability 1) whether outcome $\psi_1(x)$ or $\psi_2(x)$ will obtain, then the paths are perfectly distinguishable \emph{in this sense}, and we say that the ancilla encodes full which-way information. If the likelihoods are a random 50--50, then the paths are indistinguishable \emph{in this sense}, and we say that the ancilla encodes no which-way information. Quantitatively, for a suitably defined measure of path distinguishability $D$, the relationship between $D$ and the visibility $V=(V_\text{max}-V_\text{min})/(V_\text{max}+V_\text{min})$ of the interference fringes is given by $D^2+V^2 \le 1$, with $D^2+V^2 = 1$ for pure states.\cite{Greenberger:1988:oo,Jaeger:1995:um,Englert:1996:km,Durr:2000:za}

\section{\label{sec:theory}Theory}

We will now describe the theory of delayed-choice quantum erasure. To make contact with our experiment, we shall consider the case of a quantum eraser based on polarization interferometry. One difference to quantum eraser discussed in Sec.~\ref{sec:quantum-erasure} is that the entanglement with the ancilla is already present before the signal photon enters the interferometer. Because this entanglement describes polarization correlations and the path through the interferometer depends on polarization, which-way information is, in this sense, encoded in the ancilla \emph{prior} to the passage through the interferometer---another cautionary tale that the notion of which-way information should not be taken too literally (see Sec.~\ref{sec:notion-which-way}).

Consider a photon described by the superposition state $\ket{\psi}=\left(\ket{H}+\ket{V}\right)\sqrt{2}$, where $\ket{H}$ and $\ket{V}$ denote, respectively, horizontal and vertical polarization states. We let the photon pass through a polarization interferometer, which splits the path based on polarization: A vertically polarized photon is transmitted while a horizontally polarized photon walks off. During the photon's passage through the interferometer, the photon state acquires a relative phase $\Delta \phi$ between the components $\ket{H}$ and $\ket{V}$,
\begin{equation}\label{eq:uyfg}
\ket{\psi'}=\frac{1}{\sqrt{2}}\left(\ket{H}+\E^{\I\Delta \phi}\ket{V}\right).
\end{equation}
We then measure the polarization of the photons emerging from the interferometer. A measurement in the $HV$ basis will distinguish the two paths through the interferometer, with equal probabilities of finding the photon horizontally or vertically polarized; this is the ``particle'' picture. If we instead perform a measurement in the $\pm 45^\circ$ (diagonal) basis with corresponding eigenstates $\ket{\pm 45^\circ} =\left(\ket{H}\pm\ket{V}\right)/\sqrt{2}$, the probabilities encoded in the state~\eqref{eq:uyfg} are 
\begin{subequations}
\begin{align}
p(+ 45^\circ)  &= \abs{ \braket{+ 45^\circ}{\psi'} }^2 = \cos^2 \frac{\Delta \phi}{2},\\
p(- 45^\circ) &= \abs{ \braket{- 45^\circ}{\psi'} }^2 = \sin^2 \frac{\Delta \phi}{2}.
\end{align}
\end{subequations}
Since these probabilities exhibit an oscillatory dependence on $\Delta \phi$, an interference pattern can be obtained by varying $\Delta \phi$.

Instead of a single photon, let us now consider a pair of photons in the entangled state
\begin{equation}\label{eq:ivhdd}
\ket{\Psi}=\frac{1}{\sqrt{2}}\left(\ket{H}\ket{H}+\ket{V}\ket{V}\right).
\end{equation}
We let one of the photons in the pair (the ``signal'') pass through the polarization interferometer, resulting in the composite state 
\begin{equation}\label{eq:biudvgbhvbdi}
\ket{\Psi'}=\frac{1}{\sqrt{2}}\left(\ket{H}\ket{H}+\E^{  \I \Delta\phi}\ket{V}\ket{V}\right).
\end{equation}
Suppose we measure the signal photon in the $\pm 45^\circ$ basis, and we measure the other photon (the ``idler'') in the $HV$ basis. Then for the state~\eqref{eq:biudvgbhvbdi} the joint probability of finding $+45^\circ$ signal polarization and horizontal idler polarization is
\begin{equation}\label{eq:2a}
p(+45^\circ, H)= \frac{1}{4}.
\end{equation}
Similarly, the joint probability of finding $+45^\circ$ signal polarization and vertical idler polarization is
\begin{equation}\label{eq:3a}
p(+45^\circ,V)= \frac{1}{4}.
\end{equation}
These joint probabilities are independent of $\Delta \phi$, and thus no interference is observed even if we correlate the result of each measurement on the idler photon with the result of the corresponding measurement on the signal photon. This may not seem surprising, since the idler measurement has revealed which-way information. However, even if we do not measure the idler photon at all, no interference is observable, regardless of the basis chosen for the measurement on the signal photon. This is readily seen by considering the reduced density matrix for the signal photon, $\op{\rho}_\text{signal} = \frac{1}{2}\ketbra{H}{H}+\frac{1}{2}\ketbra{V}{V}$,  obtained by tracing (averaging) over the idler states $\ket{H}$ and $\ket{V}$ in the composite density matrix $\op{\rho}=\ketbra{\psi'}{\psi'}$ (see pp.~175--181 of Ref.~\onlinecite{Beck:2012:az} for an accessible introduction to density matrices, including examples involving photon polarization). The reduced density matrix $\op{\rho}_S$ exhaustively encodes the statistics of all polarization measurements that can be performed on the signal photon, and since it is independent of $\Delta \phi$, no interference can be observed. This shows that, for  the interference to become unobservable, it suffices that the idler photon carries which-way information in the sense represented by the state~\eqref{eq:biudvgbhvbdi}.

Therefore, in light of our discussion in Secs.~\ref{sec:quantum-erasure} and \ref{sec:notion-which-way}, in order for a relative-phase dependence (and thus interference) to show up in the measured polarization statistics, there must not exist any which-way information that would make the two paths through the interferometer distinguishable. Since the two paths are associated with the polarization states $\ket{H}$ and $\ket{V}$ of the signal photon, we must arrange matters such that there is no measurement---whether performed on the signal photons, on the idler photons, or on both---whose outcomes would allow us to infer that the probability of finding the signal photon horizontally polarized is \emph{not} equal to the probability of finding the signal photon vertically polarized. Any bias of the probabilities away from a random 50--50 split would imply a degree of distinguishability. 

To eliminate the distinguishability, we measure not only the signal photon but also the idler photon in the $\pm 45^\circ$ basis. Then for the state~\eqref{eq:biudvgbhvbdi} the joint probability of finding $+45^\circ$ polarization for both the signal and idler photon is
\begin{equation}\label{eq:2}
p(+45^\circ, +45^\circ) = \frac{1}{2}\cos^2 \frac{\Delta\phi}{2},
\end{equation}
and similarly the joint probability of finding $+45^\circ$ signal polarization and $-45^\circ$ idler polarization is
\begin{equation}\label{eq:3}
p(+45^\circ , -45^\circ) = \frac{1}{2}\sin^2 \frac{\Delta\phi}{2}.
\end{equation}
Note that these two probabilities sum to $\frac{1}{2}$ for all possible values of $\Delta \phi$: Eqs.~\eqref{eq:2} and \eqref{eq:3} represent two out-of-phase interference patterns whose sum is just the flat, $\Delta \phi$-independent no-interference pattern. 

It follows that unless the erasure measurement is actually carried out on the idler photon and the result is correlated with the result of the measurement on the signal photon, no interference pattern can be observed. To put it another way, we can never observe interference merely by looking at the results of the polarization measurements on the signal photons alone, as is already evident from the reduced density matrix, $\op{\rho}_\text{signal} = \frac{1}{2}\ketbra{H}{H}+\frac{1}{2}\ketbra{V}{V}$. Those statistics never change, no matter what action we take on the idler beam. This is just another expression of the quantum-mechanical no-signaling principle: We cannot influence the measurement statistics on one system by measuring its entangled partner. To reveal interference we must tap into the polarization correlations built into the composite state and correlate the results of two separate measurements, one performed on the signal photon and the other on the idler photon. The relative time order of these two measurements does not matter---the probabilities~\eqref{eq:2} and \eqref{eq:3} do not depend on whether we first measure the signal or the idler---but the choice between interference and path information is only open until both measurements have been carried out.

\section{\label{sec:experiment}Experiment}

\begin{figure}
\includegraphics[scale=1]{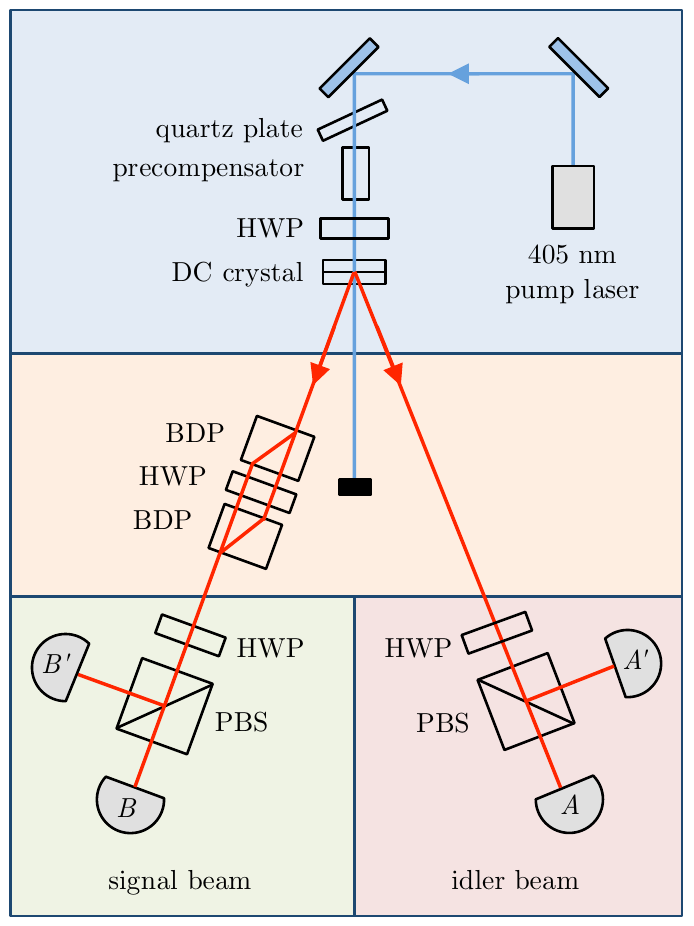}
\caption{\label{fig:setup}(Color online) Experimental arrangement of our quantum eraser based on polarization-entangled photons.  The boxes delineate the different main parts of the experiment (photon production, interferometer, and polarization measurement). Entangled 810-nm photon pairs are produced by a downconversion (DC) crystal pumped by a 405-nm diode laser. A $\unit[10 \times 10 \times 0.5]{mm}$ quartz plate adjusts the relative phase between the components in the entangled state, while a $\unit[5 \times 5 \times 5.58]{mm}$ precompensation quartz crystal improves the quality of the entanglement. The signal photon in the pair traverses an interferometer consisting of two beam-displacing prisms (BDP) and a half-wave plate (HWP). To measure photon polarization in each arm, photons pass through a half-wave plate (to adjust the measurement basis) and a polarizing beamsplitter (PBS) before being captured by fiber-coupling lenses (labeled $B$ and $B'$ in the signal arm and $A$ and $A'$ in the idler arm) and transmitted via fiber-optic cables to single-photon counting modules (not shown). A delay of the erasure measurement is implemented alternatively by elongating the optical path after the downconversion crystal in the idler beam (see Fig.~3) or by increasing the length of the fiber-optic cables in the idler arm.}
\end{figure}

The experimental arrangement of our quantum eraser is shown schematically in Fig.~2. As mentioned in the Introduction, apart from the addition of a delay stage our experimental setup is nearly identical to the quantum-eraser experiment described by Gogo \emph{et al.} in Ref.~\onlinecite{Gogo:2005:oo}. In particular, the core parts---the downconversion process, the interferometer, polarization manipulation and measurement, and photon detection---use the equipment and techniques used by Beck \emph{et al.} in a series of experiments,\cite{Thorn:2004:za,Gogo:2005:oo,Carlson:2006:za,Beck:2007:uu,Beck:2012:az,Beck:website} and we refer the reader to these references for additional details (see especially the website maintained by Beck\cite{Beck:website} for a comprehensive parts list). To make the present paper self-contained, we will nevertheless include brief descriptions of every part of our setup. The main difference between our experiment and the quantum eraser of Ref.~\onlinecite{Gogo:2005:oo} is the use of two alternative delay stages, which are described in Sec.~\ref{sec:delay-stage}.

\subsection{Photon source}

In our experiment, entangled photon pairs are produced through spontaneous parametric type-I downconversion using a pair of stacked, 0.5-mm-thick BBO crystals pumped by a 405-nm, 150-mW laser diode. The optic axes of the two crystals are oriented at $90^\circ$ with respect to each other. One crystal produces pairs of vertically polarized 810-nm photons, while the other crystal produces pairs of horizontally polarized 810-nm photons. Using a half-wave plate (HWP), the pump polarization is adjusted to equally pump both crystals. The thinness of the crystals and their close stacking means that, from the vantage point of subsequent optical elements, it is impossible to resolve in which of the two crystals a downconverted pair of photons was produced. This results in a polarization-entangled state,
\begin{equation}\label{eq:bellst}
\ket{\Psi} = \frac{1}{\sqrt{2}} \left( \ket{H}\ket{H} + \E^{\I \alpha} \ket{V}\ket{V}\right).
\end{equation}
In practice, the two components $\ket{H}\ket{H}$ and $\ket{V}\ket{V}$ are typically not perfectly indistinguishable, leading to a degradation of the entanglement. One cause is the temporal walkoff of the two orthogonal pump directions inside the downconversion crystal arising from the birefringence of the crystal. We precompensate for this walkoff by inserting a $\unit[5 \times 5 \times 5.58]{mm}$ quartz crystal (cut with its optic axis perpendicular to the direction of propagation) upstream from the downconversion crystal. We zero the relative phase $\alpha$ in the state~\eqref{eq:bellst} using an X-cut, $\unit[10 \times 10 \times 0.5]{mm}$ quartz plate mounted on a rotation stage and placed before the precompensator. To verify that our downconversion source indeed produces photons in a polarization-entangled state, we perform a test of Bell's inequalities\cite{Dehlinger:2002:aa,Dehlinger:2002:tt} and find $S=2.523 \pm 0.005$, which violates the classical bound $S \le 2$ for local realistic theories by over 100 standard deviations.

\subsection{Polarization interferometer and photon measurement}

Downconverted photon pairs emerge from the downconversion crystal at a relative angle of $6^\circ$ to each other. The signal photon then passes through two calcite beam-displacing prisms (BDPs). When the photon is incident on the first BDP, its vertically polarized component is transmitted while the horizontally polarized component is displaced by \unit[4.0]{mm}. A half-wave plate oriented at $45^\circ$ and a second BDP bring the two components spatially back together. The difference in path length between the two arms of the interferometer is adjusted by tilting the second BDP using a motorized actuator with sub-micron resolution. The tilt introduces a relative phase $\Delta \phi$ between the two components $\ket{H}$ and $\ket{V}$ in the input state $\ket{\Psi} =  \left( \ket{H}\ket{H} + \ket{V}\ket{V}\right)/\sqrt{2}$, resulting in the state given in Eq.~\eqref{eq:biudvgbhvbdi}, $\ket{\Psi'}=\left(\ket{H}\ket{H}+\E^{  \I \Delta\phi}\ket{V}\ket{V}\right)/\sqrt{2}$, where $\Delta \phi$ is proportional to the tilt of the BDP.

Past the second BDP, the polarization of the signal photon is measured using a polarization analyzer consisting of a HWP and a polarizing beamsplitter (PBS).  We orient the fast axis of the HWP at $22.5^\circ$ from the vertical, corresponding to a (subsequent) polarization measurement in the $\pm 45^\circ$ basis. Detecting a photon at output $B$ (see Fig.~2) represents a $+45^\circ$-polarized photon, while detection at $B'$ represents a $-45^\circ$-polarized photon. The polarization of the idler photons is measured by a second polarization analyzer identical to the one used in the signal beam. The setting of the HWP determines whether which-path information is erased on the idler side. Since erasure corresponds to measuring the idler photon in the $\pm 45^\circ$ basis (see Sec.~\ref{sec:theory}), for the eraser setting we orient the HWP at $22.5^\circ$ from the vertical such that the subsequent photon detection projects the state of the idler photon onto one of the diagonal states $\ket{\pm 45^\circ} =\left(\ket{H}\pm\ket{V}\right)/\sqrt{2}$. To obtain which-way information instead, the HWP is set to $0^\circ$, corresponding to a measurement in the $HV$ basis. 

Photons emerging from the output ports of the beamsplitters are captured by fiber-coupling converging lenses, fed into into multimode fiber-optic cables, and registered by single-photon counting modules (SPCMs). The SPCMs are based on silicon avalanche photodiodes, with a photon detection efficiency of about 30\% at the relevant wavelength of \unit[810]{nm}. Stray photons are removed by 780-nm long-pass filters inserted in front of the inputs of the SPCMs. In order to ensure that only pairs of single photons arising from downconversion events are detected, we register photons in coincidence between the signal and idler beams. This coincident detection also automatically guarantees the conditionalization of the results of the signal measurements on the results of the idler measurements; as discussed in Sec.~\ref {sec:theory}, this conditionalization is essential for obtaining interference patterns in a quantum eraser. Coincidence counting is performed by a field-programmable gate array (FPGA) implemented on an Altera DE2 development and education board.\cite{Lord:noyear:uu} The coincidence-time resolution $\tau_c$ of the Altera DE2 board has been measured previously and found to be consistently between 7 and 8\,ns.\cite{Lord:noyear:uu} Data are transmitted from the FPGA to a PC via an RS\,232 serial interface and displayed using Lab\textsc{view} software.

\subsection{\label{sec:delay-stage}Delay stage}

To implement a delayed-choice version of the quantum-eraser experiment, we delay the erasure measurement on the idler until the signal photons have been detected. The downconversion process leads to near-simultaneous emission of two photons within a time window on the femtosecond scale. Therefore, assuming equal optical path lengths of the signal and idler arms, in principle, a delay on the order of femtoseconds would suffice to ensure delayed erasure. However, in practice the delay time needs to be made substantially longer to take into account the limited time resolution of the SPCMs. The width $\tau_c$ of the coincidence window is a good measure for how far apart in time two photons originating from the same downconversion event may be registered. Therefore, we consider a delay time $\tau_d$ close to $\tau_c$ as sufficient for the erasure measurement to qualify as delayed. In our experiment, the erasure measurement involves three spatially separated parts. The choice of measurement basis---either $HV$ to yield which-way information, or $\pm 45^\circ$ to implement erasure---is made at the HWP in the idler beam. Subsequently, the beam is split at the PBS. Finally, the photons emerging from the two output ports of the PBS travel through fiber-optic cables and are counted at the SPCM. Where along this chain should we insert the delay?

\begin{figure}
\includegraphics[scale=1]{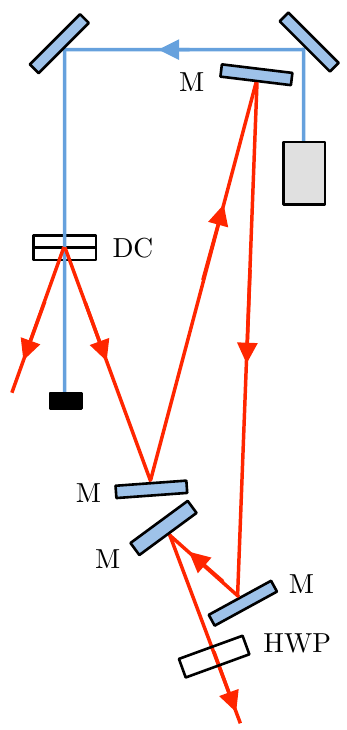}
\caption{\label{fig:mirrors}(Color online) Schematic of the free-space delay in the idler arm. Only the relevant components are shown. Four mirrors (M) are used to elongate the optical path traveled by the idler photon between the downconversion crystal (DC) and the half-wave plate (HWP) selecting the measurement basis.}
\end{figure}

Arguably, the conceptually most straightforward way is to increase the optical distance traveled by the idler photon before it reaches any of these three parts, i.e., before it passes through the HWP. In this way, no action related to the ultimate measurement (whether the action is coherent or decoherent, reversible or irreversible) is taken on the photon until after the signal photon has already been recorded by the SPCMs. We implement such a \emph{free-space delay} by inserting four mirrors into the setup that bounce the idler beam back and forth before it is allowed to reach the HWP, as illustrated in Fig.~3. We use dielectric mirrors optimized for a wavelength range of \unit[750--1100]{nm}, which includes the \unit[810]{nm} wavelength of the downconverted photons. Since a dielectric mirror introduces a phase shift between the $s$- and $p$-polarized components of the incident light, the linear polarization states used in our experiment suffer a certain amount of change upon reflection from each mirror. To minimize this undesired effect, we arrange the mirrors such that angle of incidence was as small as possible.

The achievable additional optical distance $\Delta L$ is limited by the size of the optical table (in our case, $6' \times 4'$), the geometry of the experimental arrangement, the accumulated polarization change caused by the mirrors, and beam spreading. With the mirror arrangement shown in Fig.~3, we achieve $\Delta L = \unit[2.0]{m}$, corresponding to a delay time of $\tau_d=\unit[6.7]{ns}$. This value is quite close to the width $\tau_c \approx \unit[8]{ns}$ of the coincidence window. Indeed, when the free-space delay is introduced into our setup, we lose essentially all coincidence counts, indicating that we have successfully moved outside of the coincidence window and that therefore the achieved delay time is sufficient to implement a delayed-choice experiment. To compensate for the free-space delay and restore the coincidence counts, we lengthen the path of the electrical signals going from the SPCMs in the signal arm to the FPGA. In this way, photon pairs produced by the same downconversion event will again be registered as coincident by the FPGA.

It is not actually necessary to insert the delay prior to the idler photons' reaching the HWP, because the actions of both the HWP and the PBS are coherent. The erasure itself occurs only once the idler photon is registered at the SPCM, when the which-way information is irretrievably destroyed and the information necessary to construct an interference pattern is produced. It follows that we may insert the delay anywhere upstream from the SPCMs. We realize this second version of a delay by lengthening the travel time of the idler photons downstream from the PBS, by increasing the length of the fiber-optic cable connecting the fiber-coupling lenses to the filters preceding the SPCMs. We call this configuration the \emph{fiber delay}. While in the signal arm we maintain the original length of $\unit[1.0]{m}$ for the fiber-optic cables, in the idler arm we use $\unit[5.0]{m}$-long cables instead. Since the speed of light inside the optical fiber is about $\frac{2}{3}c$, in this way we achieve a delay of about \unit[20]{ns}, which is almost three times larger than the width of the coincidence window.

\subsection{Experimental procedure}

For a given setting of the HWP in the idler beam, we record the coincidence counts $N_{AB}$ (between detectors $A$ and $B$) and $N_{A'B}$ (between $A'$ and $B$). If the idler HWP is set to erasure ($22.5^\circ$), then $N_{AB}$ represents the number of photons detected in coincidence for which both the signal and idler photon were found with $+45^\circ$ polarization. Similarly, $N_{A'B}$ represents signal photons with $+45^\circ$ polarization and idler photons with $-45^\circ$ polarization. These coincidence counts correspond to the conditional probabilities~\eqref{eq:2} and \eqref{eq:3} that we need to measure to observe interference. If the idler HWP is set to the which-way setting ($0^\circ$), then $N_{AB}$ represents signal photons with $+45^\circ$ polarization and horizontally polarized idler photons, while $N_{A'B}$ represents signal photons with $+45^\circ$ polarization and vertically polarized idler photons. These coincidence counts correspond to the conditional probabilities~\eqref{eq:2a} and \eqref{eq:3a}.

In our experiment, we first (roughly) equalize the path lengths through the interferometer by setting the idler HWP to the erasure setting ($22.5^\circ$) and adjusting the tilt of the second BDP until the visibility of the interference fringes is maximized. In a given run of the experiment, we set the idler HWP such that the coincidence counts $N_{AB}$ and $N_{A'B}$ reveal either which-way information (HWP set to $0^\circ$) or erasure information (HWP set to $22.5^\circ$). We then move the actuator in increments of about $\unit[0.44]{\mu m}$ to change the difference in path lengths between the two arms of the interferometer. Note that this value for the actuator increment is not equal to the change in linear path difference inside the interferometer, because the change in relative path length is produced by a tilt of the BDP.  At each actuator position, the coincidence counts $N_{AB}$ and $N_{A'B}$ are collected over a 5-s time interval. Finally, we graph the coincidence counts $N_{AB}$ and $N_{A'B}$ as a function of the actuator position. 

\section{\label{sec:results}Results}

\subsection{\label{sec:quantum-erasure-with}Quantum erasure with and without delay}

\begin{figure*}
\begin{flushleft}
{\small (a)}
\end{flushleft}

\includegraphics[scale=0.85]{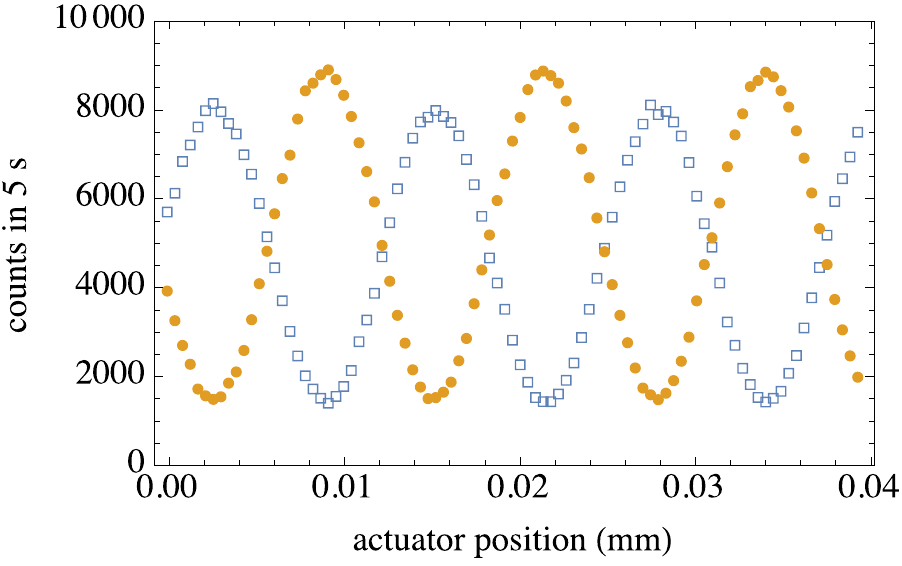} \quad
\includegraphics[scale=0.85]{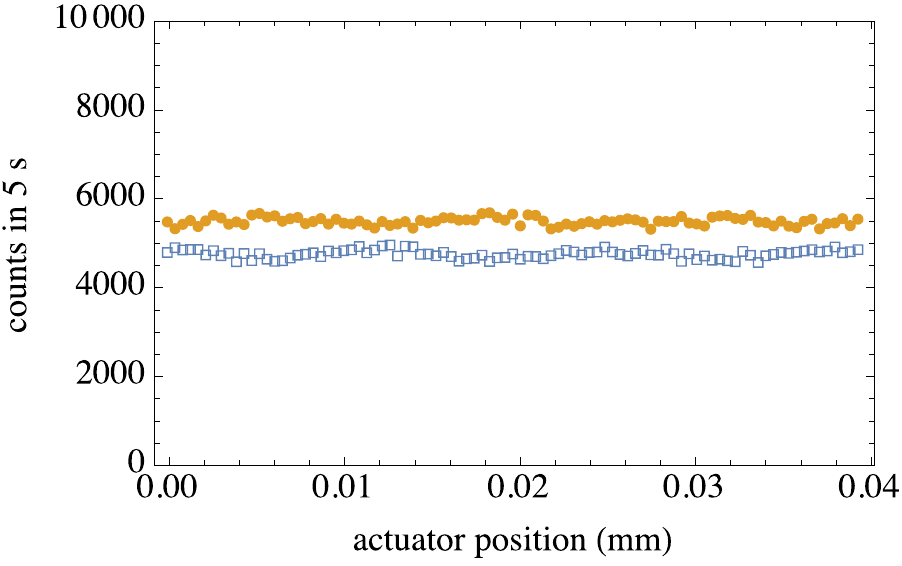}

\vspace{-.6cm}

\begin{flushleft}
{\small (b)}
\end{flushleft}

\includegraphics[scale=0.85]{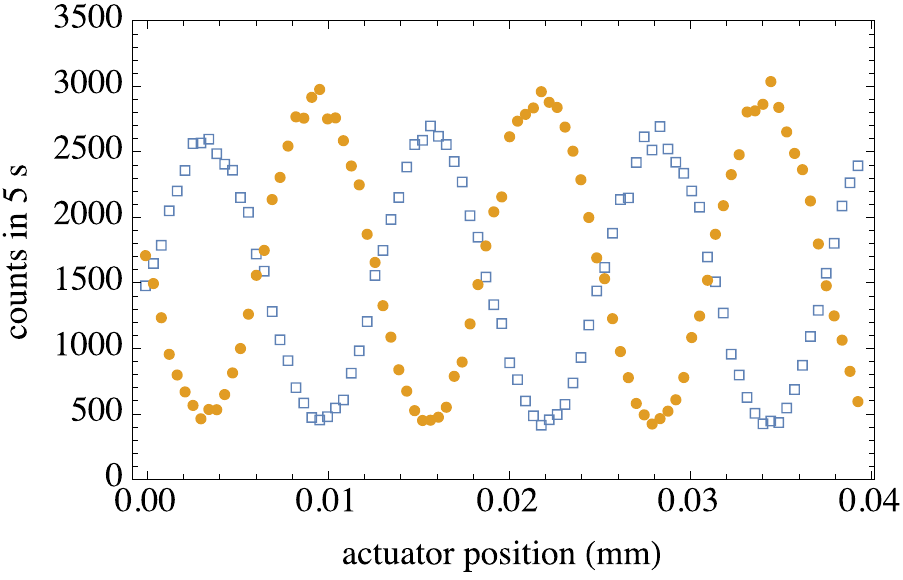}\quad
\includegraphics[scale=0.85]{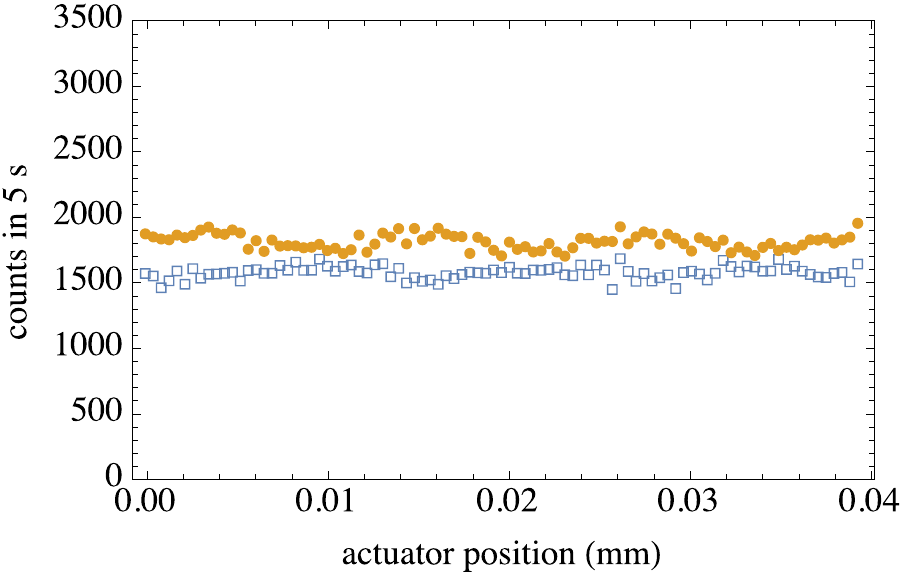}

\vspace{-.6cm}

\begin{flushleft}
{\small (c)}
\end{flushleft}

\includegraphics[scale=0.85]{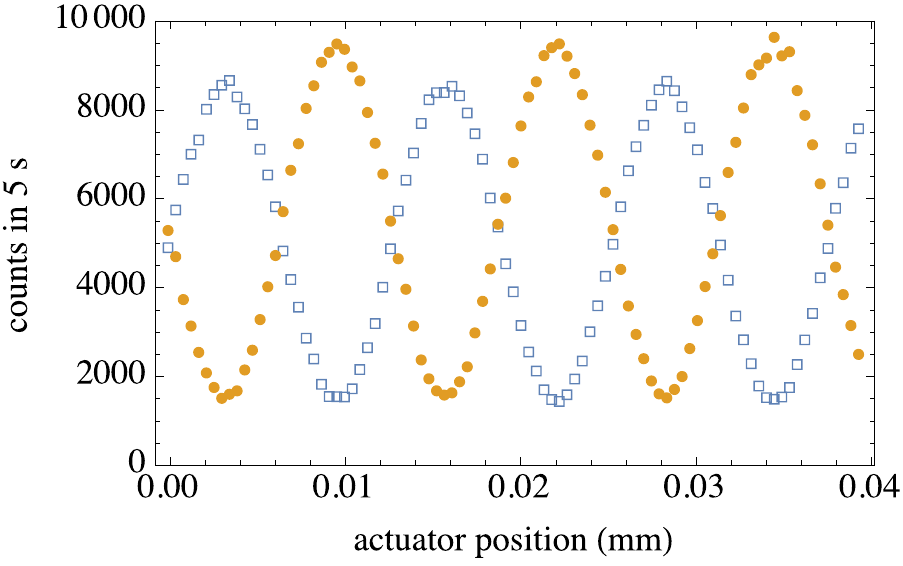}\quad
\includegraphics[scale=0.85]{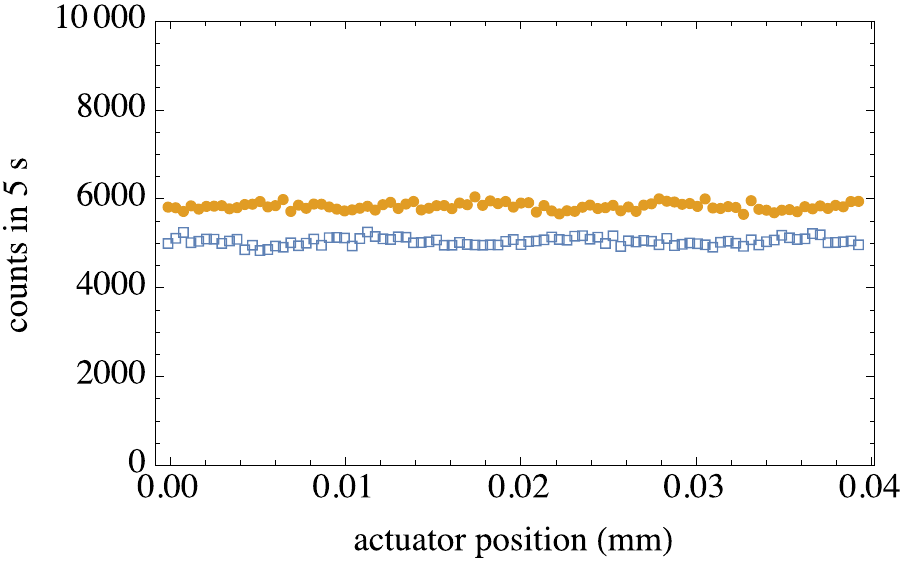}

\caption{\label{fig:results}(Color online) Coincidence counts $N_{AB}$ ($\square$) and $N_{A'B}$ ($\bullet$) as a function of the position of the actuator that adjusts the path-length difference between the two arms of the interferometer. The data in the left column are obtained when the which-way information is erased, revealing interference. The data in the right column show the absence of interference observed when the which-way information is not erased. The rows denote: (a) Without delay. (b) Free-space delay. (c) Fiber delay.}
\end{figure*}

Figure~4 shows the coincidence counts $N_{AB}$ and $N_{A'B}$ (per \unit[5]{s}) as a function of the position of the actuator (with the start position set to zero) separately for the erasure and the which-way settings. Both without and without delayed choice, we observe interference fringes with a visibility between 72\% and 75\% when the idler HWP is set to $22.5^\circ$ (erasure). The phase shift of $180^\circ$ between the two interference patterns represented by $N_{AB}$ and $N_{A'B}$ is clearly seen. Note the significantly lower count rates with the free-space delay (Fig.~4\emph{b}). We attribute this effect to the presence of the additional mirrors used to elongate the optical path, leading to a widening of the idler beam during its travel between the mirrors and thus to a reduction of the number of idler photons reaching the collection lenses. Indeed, we find the singles counts in the idler arm to be about seven times smaller than in the signal beam, while in the absence of the mirrors the singles counts in both arms are within 15\% of each other. We also find that the spacing of the interference fringes as a function of the actuator position is constant. Thus, we can conclude that, over the actuator's tilt range used in the experiment, a change in the position of the actuator translates into a proportional change of the relative path length inside the interferometer. The overall difference in the maximum count rates between the $N_{AB}$ and $N_{A'B}$ coincidences are due to a small imbalance in the efficiency between the $A$ and $A'$ detectors in capturing photons emerging from the two outputs of the PBS.

\begin{figure}

\begin{flushleft}
{\small \emph{(a)}}
\end{flushleft}


\includegraphics[scale=0.85]{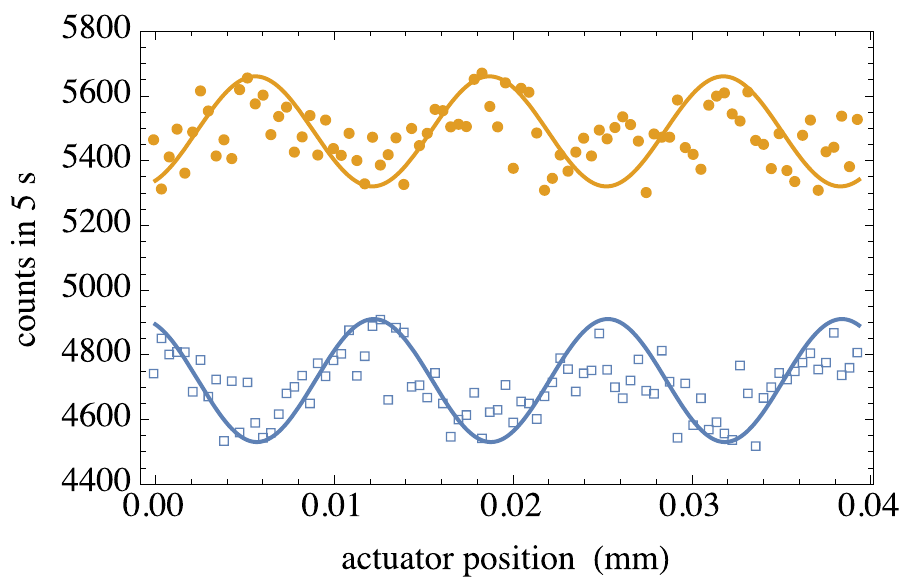}

\vspace{-.6cm}

\begin{flushleft}
{\small \emph{(b)}}
\end{flushleft}
 
\includegraphics[scale=0.85]{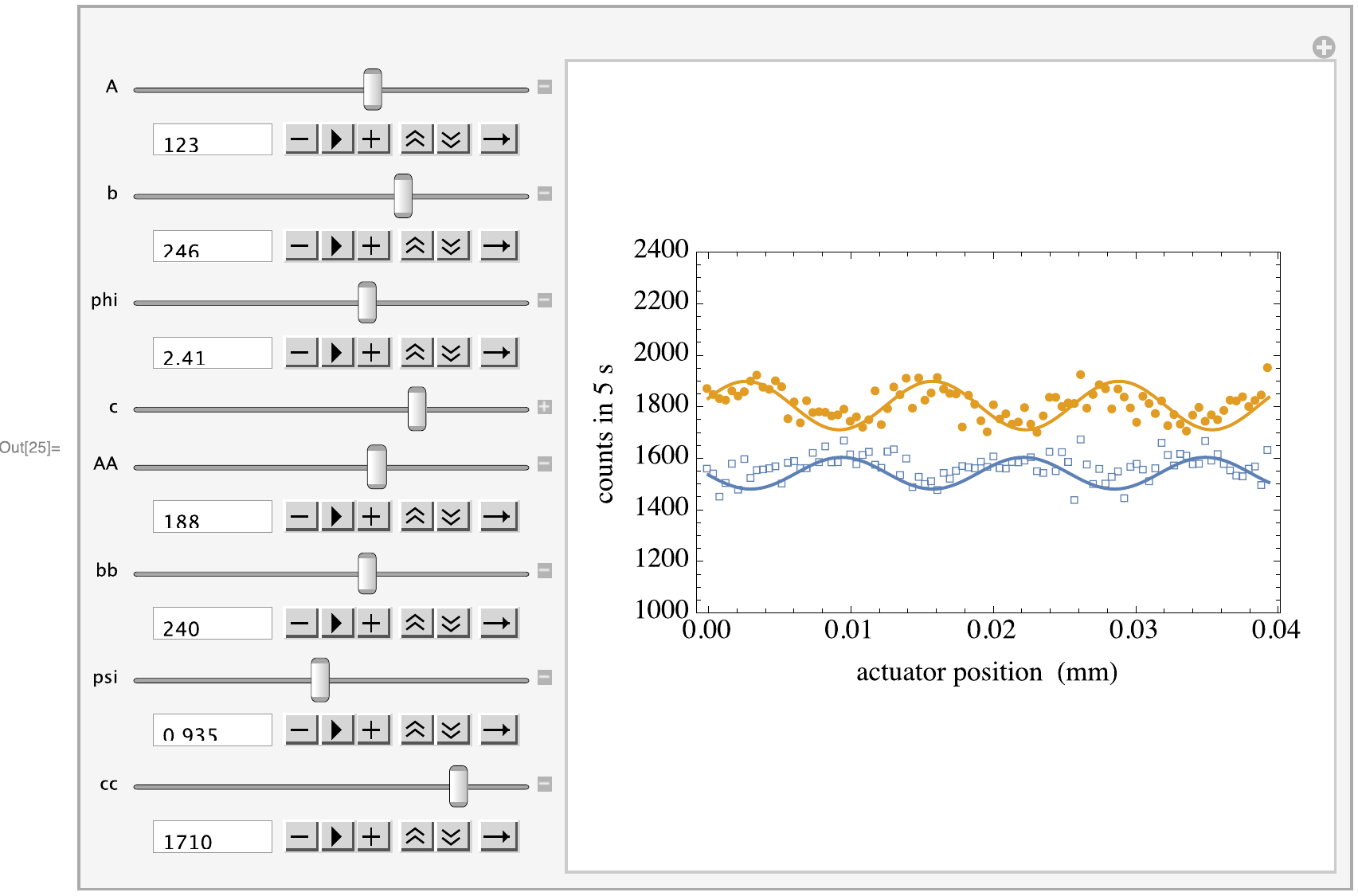}

\vspace{-.6cm}

\begin{flushleft}
{\small \emph{(c)}}
\end{flushleft}

\includegraphics[scale=0.85]{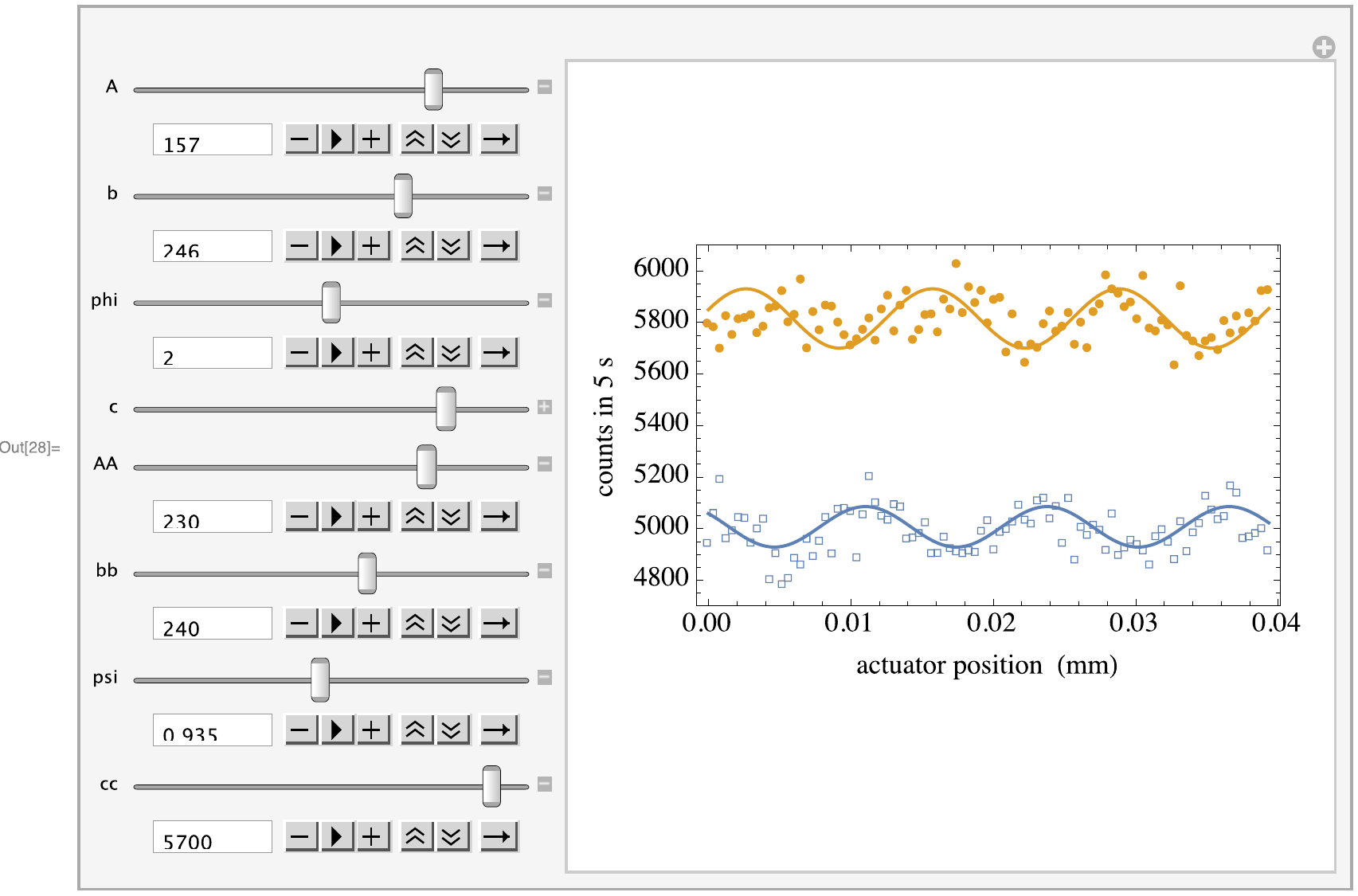}

\caption{\label{fig:residuals}(Color online) Residual variations in the coincidence counts $N_{AB}$ ($\square$) and $N_{A'B}$ ($\bullet$) for the which-way setting. A sine-squared fit has been applied by hand. The periodic variations match the period of the interference fringes, indicating remaining phase information. The fluctuations about those variations are dominated by Poisson counting noise. As in Fig.~4, the rows indicate: (a) Without delay. (b) Free-space delay. (c) Fiber delay.}
\end{figure}

By setting the idler HWP to $0^\circ$, the interference fringes essentially disappear, as seen in Fig.~4. We observe small residual fluctuations in the count rates. Figure~5 takes an enlarged look at these fluctuations (note the reduced range of the vertical axis), with a sine-squared fit applied by hand. A sinusoidal variation of the count rates can be discerned, and the period is found to match the period of the interference fringes shown in the left column of Fig.~4. These observations indicate that the sinusoidal variations represent low-visibility interference fringes arising from a small amount of phase information that has remained available. This effect is likely caused by imperfect state preparation and small inaccuracies in the idler wave-plate settings. The latter cause is also suggested by the observed relative phase shift of $180^\circ$ between the residual fluctuations and the high-visibility fringes. We attribute this phase shift to a small overshoot in the rotation of the idler wave plate in our experiment for the which-way setting (idler HWP at $0^\circ$), such that the state component that on the erasure setting (idler HWP to $22.5^\circ$) had previously emerged from a particular output of the PBS is now directed to the opposite output. Thus, the fringes seen in the $N_{AB}$ counts on the erasure setting subsequently appear, in reduced form, in the $N_{A'B}$ counts on the which-way setting, such that the roles of the $A$ and $A'$ detectors has become effectively swapped. 

Figure~5 shows that the count rates also fluctuate about their overall sinusoidal variation. Likely causes include Poisson noise of the photon-counting statistics, phase fluctuations in the interferometer, and accidental coincidences resulting from the finite width of the coincidence window. For our data, we find the expected variation $\langle N \rangle^{1/2}$ due to Poisson counting statistics to be only 10\%--15\% lower than the standard deviation of each sample. Given that the standard deviation is also influenced by the sinusoidal variations, this suggests that Poisson noise is the dominant source of the observed nonperiodic fluctuations.

\subsection{The role of entanglement}

As shown by Gogo \emph{et al.},\cite{Gogo:2005:oo} the essential phenomenology of the quantum eraser does not require an entangled input state [as in Eq.~\eqref{eq:ivhdd}] but can also be replicated by the mixed input state 
\begin{align}\label{eq:kdjgvkj}
\op{\rho}&=\frac{1}{2}\ketbra{+45^\circ,+45^\circ}{+45^\circ,+45^\circ}\notag \\&\quad +\frac{1}{2}\ketbra{-45^\circ,-45^\circ}{-45^\circ,-45^\circ},
\end{align}
where $\ket{\pm 45^\circ,\pm 45^\circ} \equiv \ket{\pm 45^\circ}\ket{\pm 45^\circ}$. This state describes a situation in which each individual signal photon is in one of the two states $\ket{\pm 45^\circ}$ prior to entering the interferometer. Both of these states lead to interference patterns when the polarization of the signal photon is subsequently measured in the $\pm 45^\circ$ basis. However, just as for the entangled state~\eqref{eq:ivhdd}, the two interference patterns are out of phase with each other. Thus, in any given run of the experiment, the mixed state~\eqref{eq:kdjgvkj} will produce statistics that do not show interference. If the polarization of the idler photon is measured in the $\pm 45^\circ$ basis, then we obtain the information necessary to separate out these two interference patterns. If we instead measure the idler photon in the $HV$ basis, then this measurement yields no information about whether the signal photon was prepared in the state $\ket{+45^\circ}$ or in the state $\ket{-45^\circ}$. Therefore, we cannot distinguish the out-of-phase interference patterns, and even correlating the signal and idler measurements will not reveal interference. 

The ability of the mixed state~\eqref{eq:kdjgvkj} to mimic quantum-eraser behavior raises the question of how one may experimentally distinguish this state from the entangled state~\eqref{eq:ivhdd}.  One approach, realized by Gogo \emph{et al.},\cite{Gogo:2005:oo} is to insert a beam block into one of the paths through the interferometer (say, into the path corresponding to vertically polarized signal photons), which effectively amounts to a polarization measurement in the $HV$ basis \emph{inside} the interferometer. When both signal and idler subsequently measured in the $HV$ basis, the coincidence count rate $N_{HH}$ (corresponding to finding both signal photon and idler photon horizontally polarized) remains unchanged for the entangled state while it is reduced by 50\% for the mixed state.\cite{Gogo:2005:oo}

Here we use a different method for distinguishing the entangled and mixed states. Suppose the downconversion crystal produces photon pairs in the mixed state
\begin{equation}\label{eq:kdigyjgvkj}
\op{\rho}=\frac{1}{2}\ketbra{H,H}{H,H}+\frac{1}{2}\ketbra{V,V}{V,V}.
\end{equation}
Then state~\eqref{eq:kdjgvkj} can be produced by rotating the polarization of the signal and idler photons by $45^\circ$ after they emerge from the downconversion crystal (as done experimentally in Ref.~\onlinecite{Gogo:2005:oo}). Because the unrotated state~\eqref{eq:kdigyjgvkj} can never exhibit interference, it follows that these joint rotations are critical for observing interference and erasure behavior. By contrast, the entangled state~\eqref{eq:ivhdd} is invariant under the same rotations, which should therefore not affect the behavior of the quantum eraser. We confirm this prediction by repeating our experiment, using the entangled state~\eqref{eq:ivhdd} but this time inserting HWPs into the signal and idler paths immediately following the downconversion crystal. The purpose of these HWPs is to rotate the photon polarizations, just as one would do when creating state~\eqref{eq:kdjgvkj} from state~\eqref{eq:kdigyjgvkj}. In each run of the experiment, both HWPs are jointly set to a new orientation (in the range 0--$30^\circ$) and our previous erasure and which-way settings for the polarization analysis are used. The resulting coincidence counts are found to be consistent with the no-rotation results reported in Sec.~\ref{sec:quantum-erasure-with} through each angle of the HWPs. Thus, for the entangled state we find no significant effect of the state rotations on our ability to observe both interference and which-way behavior.

Given that quantum-eraser behavior can be mimicked using mixed states, one might wonder about the motivation for using entangled states in this experiment, especially given that such states can be difficult to produce. We suggest three motivations. First, since the correlations represented by the mixed state are purely classical and thus hardly mysterious, the corresponding erasure process facilitated by these correlations is similarly unsurprising. Second, one of the driving ideas that underlies the quantum-eraser model, as well as related phenomena such as quantum decoherence, is that loss of coherence can arise purely from entanglement---i.e., from quantum correlations---with a which-way marker, rather than requiring a physical disturbance or irreducible state reduction. Therefore, in order to meet the features and conditions of quantum erasure commonly put forward in the literature (see, e.g., the discussion by Kwiat \emph{et al.}\cite{Kwiat:1994:za}), it becomes desirable to use entangled states when implementing a quantum eraser. A third motivation for using entangled states is practical. A setup like ours will typically be used for a series of different quantum-optics experiments, including experiments in which entangled states are already produced. Creating the mixed state~\eqref{eq:kdjgvkj}, on the other hand, would require additional and relatively expensive equipment, such as the liquid crystal variable retarder used in Ref.~\onlinecite{Gogo:2005:oo}.

\section{\label{sec:discussion}Discussion}

Implementation of a delayed-choice experiment in the strict sense requires closure of several loopholes familiar from tests of Bell's inequalities.\cite{Larsson:2014:ja} For example, any action pertaining to the erasure, including the choice of whether to erase the which-way information, must be relativistically separated from the passage of the signal photon through the interferometer, and the choice itself must be made by a genuinely random process. Such an experiment has been done\cite{Ma:2013:un} using ultrafast switching, large spacelike separations between the interferometer and the site of the erasure measurement, and a quantum random number generator determining the choice of measurement setting. Clearly, experiments of this kind are outside of the scope of a setup such as ours. 

Our experiment, however, comes quite close to the ideal. To be sure, the setting of the wave plate in the idler beam, which determines whether the ultimate polarization measurement will yield which-way or erasure information, is chosen at the beginning of the experiment, well before the signal photon enters the interferometer. But what the free-space delay used in our experiment does achieve is to delay the passage of the idler photon through the wave plate until after the signal photon has been registered. Therefore, all erasure-related actions on the idler photon are genuinely delayed. The alternative implementation of a delay we have used---in which the delay is introduced only past the polarizing beamsplitter in the idler beam---also properly implements a delayed erasure, because the erasure is only produced once the photon has been detected at the photon counter. 

The two different methods we have used for delaying the erasure measurement each have their advantages and drawbacks. The free-space delay is pedagogically and conceptually simple but has practical shortcomings. The achievable delay time is rather limited (in our experiment, we barely reach the width of the coincidence window), the mirrors cause a small, undesired change of the polarization state, and the longer travel distance leads to a beam broadening that significantly reduces the number of idler photons reaching the detection optics. The fiber delay does not have any of these limitations. While the achievable delay time is ultimately limited by photon losses inside the optical fiber, for the additional cable length of \unit[4.0]{m} used in our experiment we find no noticeable drop in coincidence counts compared to the situation in which the fiber delay is absent. Therefore, one could realize significantly longer delay times than those of our experiment. However, since the delay time achieved in our experiment is already much longer than the coincidence window, doing so is unlikely to be useful. On the conceptual side, it may require some effort to convince a student of the fact that the elongated fiber indeed implements a physically proper delay of the erasure. Yet, the pedagogical value of this exercise should not be underestimated.

In our experiment, we observe no change in the results of our quantum-eraser experiment when the erasure of the which-way information is delayed until after the signal photon has already been detected. Both in the absence and presence of a delay, we obtain interference fringes with very similar visibility (between 72\% and 75\%) if the measurement arrangement is such that the idler photon is projected onto a polarization state that obliterates the which-way information. If we instead choose to learn the which-way information, then no fringes are seen, save for small residual fluctuations in the form of a periodic variation (representing interference fringes of very low visibility) with added Poisson counting noise. The observation that the delay does not make a difference to the measurement statistics follows from a simple but far-reaching fact of quantum theory, namely, that the time order of quantum measurements on separate systems (or degrees of freedom) is irrelevant. The two scenarios in a quantum-eraser experiment---the observation of interference indicative of a ``wave-like'' behavior of the photon, and the retrieval of which-way information suggesting a ``particle-like'' behavior---are not statements about the behavior of the signal photon. Rather, they correspond to correlating data obtained from a measurement on the signal photon---data that by itself never shows interference---with the results of two different, mutually exclusive measurements on the idler photon. The delayed-choice version of the quantum eraser makes this lesson particularly clear.

We close with a quotation from Wheeler's first paper introducing the idea of delayed-choice experiments:\cite{Wheeler:1978:az}
\begin{quote}
Not one of the seven delayed choice experiments [presented in the paper] has yet been done. There can hardly be one that the student of physics would not like to see done. 
\end{quote}
By ``student,'' Wheeler was likely referring to any physicist interested in quantum mechanics. Our experiment, however, grants Wheeler's implicit wish in its stated sense. It can be set up and carried out by undergraduate students, opening up the world of delayed-choice experiments to this audience and enriching laboratory and lecture courses alike.

\begin{acknowledgments}
The authors are indebted to M.~Beck, R.~Haskell, and T.~Lynn for equipment advice and valuable conversations, and for hosting two workshops on single-photon experiments sponsored by the Advanced Laboratory Physics Association. The authors  also thank J.\ Essick for helpful discussions. The authors are grateful to the two anonymous reviewers for prompting us to investigate more closely the residual interference fringes and the distinction between entangled and mixed states. This research was supported by the Student Summer Science Scholar program of the University of Portland and by the Foundational Questions Institute.
\end{acknowledgments}


\end{document}